\documentclass[superscriptaddress,letterpaper, amssymb,preprint, amsmath]{revtex4-1}
\usepackage[english]{babel}
\usepackage[utf8]{inputenc}
\usepackage{mathtools}
\usepackage{braket}
\usepackage{xcolor}
\usepackage[symbol]{footmisc}
\usepackage{scalerel}
\usepackage{tikz}
\usetikzlibrary{svg.path}
\definecolor{orcidlogocol}{HTML}{A6CE39}
\tikzset{
  orcidlogo/.pic={
    \fill[orcidlogocol] svg{M256,128c0,70.7-57.3,128-128,128C57.3,256,0,198.7,0,128C0,57.3,57.3,0,128,0C198.7,0,256,57.3,256,128z};
    \fill[white] svg{M86.3,186.2H70.9V79.1h15.4v48.4V186.2z}
                 svg{M108.9,79.1h41.6c39.6,0,57,28.3,57,53.6c0,27.5-21.5,53.6-56.8,53.6h-41.8V79.1z M124.3,172.4h24.5c34.9,0,42.9-26.5,42.9-39.7c0-21.5-13.7-39.7-43.7-39.7h-23.7V172.4z}
                 svg{M88.7,56.8c0,5.5-4.5,10.1-10.1,10.1c-5.6,0-10.1-4.6-10.1-10.1c0-5.6,4.5-10.1,10.1-10.1C84.2,46.7,88.7,51.3,88.7,56.8z};}}
\newcommand\orcidicon[1]{\href{https://orcid.org/#1}{\mbox{\scalerel*{
\begin{tikzpicture}[yscale=-1,transform shape]
\pic{orcidlogo};
\end{tikzpicture}
}{|}}}}
\usepackage[colorlinks,citecolor=red,urlcolor=blue,bookmarks=false,hypertexnames=true]{hyperref} 

\newcommand{\kwv}{\mathbf{k}}
\newcommand{\qwv}{\mathbf{q}}
\newcommand{\phn}{N_{\nu \qwv}}
\newcommand{\scm}{\Theta_{m\kwv \\ m'\kwv'}}
\newcommand{\fkm}{f_{m\kwv}}
\newcommand{\vcm}{$\rm V \, cm^{-1}$}
\begin{document}

\title{Electronic noise of warm electrons in semiconductors from first-principles}

\author{Alexander Y. Choi \orcidicon{0000-0003-2006-168X}}
\thanks{These authors contributed equally to this work.}
\affiliation{Division of Engineering and Applied Science, California Institute of Technology, Pasadena, CA, USA}

\author{Peishi S. Cheng \orcidicon{0000-0002-3513-9972}}
\thanks{These authors contributed equally to this work.}

\author{Benjamin Hatanp{\"a}{\"a} \orcidicon{0000-0002-8441-0183}}

\affiliation{Division of Engineering and Applied Science, California Institute of Technology, Pasadena, CA, USA}

\author{Austin J. Minnich \orcidicon{0000-0002-9671-9540}}
\thanks{Corresponding author: \href{mailto:aminnich@caltech.edu}{aminnich@caltech.edu}}
\affiliation{Division of Engineering and Applied Science, California Institute of Technology, Pasadena, CA, USA}

\date{\today} 

\begin{abstract}
The \textit{ab-initio} theory of low-field electronic transport properties such as carrier mobility in semiconductors is well-established. However, an equivalent treatment of electronic fluctuations about a non-equilibrium steady state, which are readily probed experimentally, remains less explored. Here, we report a first-principles theory of electronic noise for warm electrons in semiconductors. In contrast with typical numerical methods used for electronic noise, no adjustable parameters are required in the present formalism, with the electronic band structure and scattering rates calculated from first-principles. We demonstrate the utility of our approach by applying it to GaAs and show that spectral features in AC transport properties and noise originate from the disparate time scales of momentum and energy relaxation, despite the dominance of optical phonon scattering. Our formalism enables a parameter-free approach to probe the microscopic transport processes that give rise to electronic noise in semiconductors.

\end{abstract}

\maketitle
\section{Introduction}\label{sec:Introduction}
Charge transport in semiconductors is a topic of fundamental and practical interest with a well-established theoretical foundation \cite{Ziman_1960,Ferry_2000}. In many cases, a sufficient understanding of the relevant physics at both low and high fields can be achieved using the Boltzmann equation with semi-empirical scattering rates \cite{Lundstrom_2000,Chen_2005,Mahan_2011, conwell_1967}. In other cases, a more precise description of the electronic transitions induced by phonons and other perturbations is required. Such a description is now possible owing to advances in electronic structure codes that enable the \textit{ab-initio} computation of the transition matrix elements given by Fermi's golden rule performed in conjunction with the numerical solution of the Boltzmann equation describing carrier dynamics \cite{Liao_2015, Bernardi_2016,Giustino_2017,Bernardi_2014}. While method development is ongoing, these calculations are now routine for various semiconductors including Si \cite{Ponce_2018,Fiorentini_2017, Li_2015}, GaAs \cite{Zhou_2016,Liu_2017,Lee_2018_b}, phosphorene \cite{Liao_2015}, and others \cite{Lee_2018_a, Jhalani_2017,Ma_2018,Singh_2015}.

In contrast, an equivalent treatment of fluctuations from a non-equilibrium steady-state is lacking, despite the experimental accessibility of electronic noise \cite{Hartnagel_2001,Bareikis_book} and its importance in applications \cite{pospi_2013}. At equilibrium, the Nyquist relation, or more generally the fluctuation-dissipation theorem, relates the electrical conductivity to the spectral noise power \cite{Kogan_1996, Callen_1952, Nyquist_1928}. Outside of equilibrium, the theorem no longer applies and the spectral noise power must be computed with another approach.

The theoretical description of fluctuations about a non-equilibrium steady-state  has a long history. In 1935, Leontovich used kinetic theory to examine velocity fluctuations of a non-equilibrium gas  \cite{Leontovich_1935}. Around 20 years later, Wannier established the definition of a diffusion coefficient for transport about a non-equilibrium steady state \cite{Wannier_1952}. Hashitsume considered a microscopic description of occupancy fluctuations about a steady distribution using the  Fokker-Planck equation with a random source term \cite{Hashitsume_1956}. In analogy with earlier works on fluctuational Maxwell equations, Kadomotsev introduced Langevin sources into the Boltzmann equation \cite{Kadomotsev_1957}. Shortly thereafter, Price derived that for spatially homogeneous fluctuations, a fluctuation-diffusion relation links Wannier's diffusion coefficient to the spectral density of current fluctuations even outside of equilibrium \cite{Price_1960}. For this reason, the non-equilibrium noise at frequencies small compared to scattering rates is known as diffusion noise. In the same year, Lax formulated a general kinetic theory of fluctuations for a Markovian system \cite{Lax_1960}. Throughout the 1960s, Gantsevich and co-workers applied Lax's kinetic theory to dilute gases for which the evolution of the one particle distribution function is governed by the linear Boltzmann Equation \cite{GGK_1969}. Their technique, termed the  ``method of moments,'' demonstrated how to compute the spectral density of current fluctuations using only the solutions of the linear Boltzmann equation. Concurrently with Gantsevich, starting from Kadomotsev's Boltzmann-Langevin equation, Kogan and Shul'man developed a Langevin treatment of the current density fluctuations \cite{Kogan_1969}. Lax, van Vliet, and Kogan and Shul'man independently confirmed that the method of moments and the Langevin approach are equivalent \cite{Lax_1966,vanVliet_1965,Kogan_1970}.

As computational resources became increasingly available, numerical implementations of the methods described above permitted computations of electronic noise for both warm ($\Delta T/ T_0 \ll 1$) and hot ($\Delta T/ T_0 \sim 1$) electrons, where $\Delta T$ is the steady-state temperature rise of the electrons and $T_0$ is the lattice temperature. Due to the lack of knowledge of the precise transition rates between electronic states,  these studies employed simplified band structures and parameterized models for scattering such as deformation potential theory for acoustic phonon scattering \cite{regianni_1983, reggiani_1997}. For example, Stanton and Wilkins obtained the Green's function of the Boltzmann equation under the single-mode relaxation time approximation, demonstrating qualitative agreement with experiment in GaAs for one \cite{Stanton_1987_1} and two \cite{Stanton_1987_2} valleys. Numerous Monte Carlo simulations reported calculations of current spectral densities in Si \cite{Gherardi_1975,Fauquembergue_1980,Ferry_1981,regianni_1983,Kuhn_1990, varani_1994}, GaAs \cite{fawcett_1969,Fauquembergue_1980,Bosi_1976}, and other semiconductors \cite{Hill_1979,Wang_2012,Starikov_2005,Rengel_2013}. These works employed various approximations such as Debye acoustic phonons, dispersionless optical phonons, and spherical approximations for electron conduction bands. With empirical knowledge of band structure parameters such as effective mass and approximate relaxation times, reasonable agreement with experiments was reported \cite{Gherardi_1975,Fauquembergue_1980,Ferry_1981,regianni_1983,Kuhn_1990,varani_1994,fawcett_1969,Fauquembergue_1980,Hill_1979,Xing_1988_1,Xing_1988_2}. More recently, these methods have been extended to heterostructures and have provided  insight into the design  of low noise devices \cite{Mateos_2008,Mateos_2009,mateos_2017}. While studies with parameterized models can provide an adequate description of the physics of interest in certain cases, they are not predictive and are restricted to materials for which empirical models of the dominant scattering mechanisms are available. It is therefore natural to consider how advances in the \textit{ab-initio} calculation of mean transport quantities \cite{Bernardi_2016,Giustino_2017} can be applied to the non-equilibrium steady state.

Here, we present an \textit{ab-initio} theory of electronic noise for warm electrons in non-degenerate semiconductors. The formalism provides the spectral noise power and AC transport quantities without any adjustable parameters. Using the method, we show that the anisotropy and spectral features of the noise power in GaAs can be explained by the disparate timescales of momentum and energy exchange with phonons, even though scattering is dominated by the inelastic polar optical phonon scattering mechanism. The formalism is easily extendable to other semiconductors of technological interest such as InP, Si, and Ge. Our method  provides  a  parameter-free view of the microscopic transport processes responsible for electronic fluctuations in semiconductors and will advance fundamental studies of carrier transport and  applications of low noise semiconductor devices. 

\section{Theory}\label{sec:Theory}

\subsection{Steady-state transport}

We begin by reviewing the \textit{ab-initio} treatment of steady-state transport using the Boltzmann equation to set the notation. Consider a non-degenerate, spatially homogeneous electron gas subject to an external electric field $\mathcal{E}$. The system is governed by the following Boltzmann equation:

 \begin{equation}
    \frac{\partial f_{m\kwv}}{\partial t} + \sum_{\gamma}\frac{e\mathcal{E}_{\gamma}}{\hbar} \, \frac{\partial \fkm}{\partial_{k_{\gamma}}} = \mathcal{I}[\fkm]
 \end{equation}

Here, $f_{m\kwv}$ is the distribution function that describes the occupancy of the electron state with wave vector $\mathbf{k}$ and band index $m$, $e$ is the fundamental charge,  $\hbar$ is the reduced Planck constant, and $\gamma = x, y, z$ indexes the crystal axes. 

The collision integral, $\mathcal{I}$,  describes the scattering rates between electronic Bloch states $m\kwv$ and $m'\kwv'$. In general, the collision integral is a nonlinear functional of the distribution function given by Fermi's Golden Rule \cite{Ziman_1960}. In the steady case, the transient term vanishes, and we denote the solution of the resulting equation as $\fkm^s$. 

In many problems, a good approximation is that the  Boltzmann equation can be linearized about an equilibrium distribution as $\fkm^s \equiv \fkm^{0} + \Delta \fkm$, where  $\Delta f_{m\kwv}$ is the change in occupation due to the electric field $\mathcal{E}$ relative to the equilibrium distribution $\fkm^{0}$. Under the non-degenerate assumption, $\fkm^0$ is well approximated by the Maxwell-Boltzmann distribution. With this substitution and retaining only terms linear in $\Delta f_{m\kwv}$, the Boltzmann equation becomes \cite{Lundstrom_2000}:

\begin{equation}\label{linbte}
    \sum_{\gamma} \bigg[ \frac{e \mathcal{E}_{\gamma}}{\hbar} \, \frac{\partial \Delta \fkm}{\partial_{k_{\gamma}}} \bigg] + \sum_{m'\kwv'} \scm \Delta f_{m'\kwv'} = - \sum_{\gamma} \frac{e \mathcal{E}_{\gamma}}{\hbar}   \, \frac{\partial \fkm^0}{\partial_{k_{\gamma}}} 
\end{equation}

\noindent where  $\scm$ is the linearized collision integral: 
 
\begin{equation}\label{matrixel}
    \scm = \frac{2 \pi}{\mathcal{N} \hbar} \sum_{m'\nu \qwv}  |g_{m \kwv,m'\kwv + \qwv}|^2  \bigg[\delta(\epsilon_{m \mathbf{k}}-\hbar \omega_{\nu \mathbf{q}}-\epsilon_{m' \kwv+\qwv}) H_{em} + \delta(\epsilon_{m\mathbf{k}} + \hbar \omega_{\nu\mathbf{q}} -\epsilon_{m'\kwv+\qwv}) H_{abs} \bigg]
\end{equation} 

Here, $g_{m \kwv,m'\kwv'}$ is the matrix element coupling electron state $m\mathbf{k}$ to another electron state $m'\mathbf{k'} = m'\kwv+\qwv$ via emission or absorption of a phonon with wave vector $\mathbf{q}$, polarization $\nu$, and occupancy $\phn$ given by the Bose distribution.  $\mathcal{N}$ is the total number of $\qwv$-points. The linearized emission and absorption weights are $H_{ems} = N_{\mathbf{q}} + 1 - f^0_{m\mathbf{k+q}}$ and $H_{abs} = N_{\mathbf{q}} + f^0_{m\mathbf{k+q}}$, respectively, if $m'\kwv' = m\kwv$, and  $H_{ems} = -(N_{\mathbf{q}} + \fkm^0)$ and $H_{abs} = -(N_{\mathbf{q}} + 1 - \fkm^0)$  if  $m'\kwv' \neq m\kwv$. Note that in Eqn. \ref{linbte} we have moved the collision integral to the left-hand side and defined Eqn.~\ref{matrixel} without the usual minus sign to simplify the following expressions.

In the present study, we  restrict the electric field to values where $\Delta \fkm \ll \fkm^0$ so that the linearization above is valid. However, in the typical \textit{ab-initio} treatment of transport, the electric field is further assumed to be small enough such that $\partial \fkm/\partial_{k_{\gamma}} \approx \partial \fkm^{0} / \partial_{k_{\gamma}}$, allowing $\Delta \fkm$ to be obtained by an iterative method with only knowledge of $\scm$ and the equilibrium distribution $\fkm^0$ \cite{Bernardi_2016}. In the present problem, the field is sufficiently large such that $\partial \Delta \fkm / \partial_{k_{\gamma}} \sim \partial \fkm^0  / \partial_{k_{\gamma}}$ and the  neglected derivative term, $\partial \Delta \fkm / \partial_{k_{\gamma}}$, must be included. This approximation was originally denoted as  the `warm electron' approximation  since the excess energy of the electrons over the thermal value can be non-zero  while remaining small on that scale \cite{conwell_1967}.

To treat the drift term numerically, we employ a finite difference approximation:

\begin{equation}
    \sum_{\gamma}\frac{e\mathcal{E}_{\gamma}}{\hbar}  \, \frac{\partial \Delta \fkm^s}{\partial_{k_{\gamma}}} \approx  \sum_{\gamma} \frac{e \mathcal{E}_{\gamma}}{\hbar} \sum_{m'\kwv'} D_{m\kwv m'\kwv',\gamma} \, \Delta f_{m'\kwv'} 
\end{equation}

\noindent where the momentum-space derivative is approximated using the finite-difference scheme given in \cite{mostofi_2008} and Eqn.~8 of Ref.~\cite{Pizzi_2020}.

With these definitions, the steady Boltzmann equation becomes:

\begin{equation}\label{lambda}
    \sum_{m'\kwv'} \Lambda_{m \kwv m'\kwv'} \Delta f_{m'\kwv'} \equiv \sum_{\gamma} \sum_{m'\kwv'}  \bigg[\frac{e \mathcal{E}_{\gamma}}{\hbar}D_{m\kwv m'\kwv',\gamma} + \scm \bigg] \Delta  f_{m'\kwv'} = \sum_{\gamma} \frac{e \mathcal{E}_{\gamma}}{k_B T} v_{m\kwv, \gamma} \fkm^0
\end{equation}

\noindent Here, we have analytically expanded the gradient of the equilibrium Boltzmann distribution on the right-hand side as $\partial \fkm^0 /\partial_{k_{\gamma}} = -(\hbar v_{m\kwv, \gamma}/k_B T) \fkm^0$, where $v_{m \kwv}$ is the group velocity and $k_B T$ is the thermal energy. $\Lambda_{m\kwv m'\kwv'}$ is defined as the relaxation operator that combines the drift and scattering operators. Equation~\ref{lambda} shows that the steady Boltzmann equation is now a system of linear equations. The solution, $\Delta \fkm$, can be written symbolically using the inverse relaxation operator:

\begin{equation}\label{steady_bte}
    \Delta \fkm =   \sum_{m'\kwv'}\Lambda_{m\kwv m'\kwv'}^{-1} \sum_{\gamma} \bigg(\frac{e \mathcal{E}_{\gamma}}{k_B T}   \bigg) v_{m'\kwv', \gamma} f_{m'\kwv'}^0
\end{equation}

Transport properties such as the electrical conductivity can be defined using the steady distribution. In particular, the linear DC conductivity $\sigma_{\alpha \beta}^{lin}$ can be expressed as:

\begin{equation}\label{lowfield_cond}
    \sigma_{\alpha \beta}^{lin} =  \frac{2e^2}{k_B T \mathcal{V}_0} \sum_{m\kwv} v_{m \kwv,\,\alpha} \sum_{m'\kwv'} (\scm^{-1} \, v_{m'\kwv', \,\beta} f_{m'\kwv'}^0)
\end{equation}

where the factor of 2 accounts for spin degeneracy and $\mathcal{V}_0$ is the supercell volume. The field is applied along the $\beta$ axis and the resulting current is measured along the $\alpha$ axis. The conductivity of Eqn.~\ref{lowfield_cond}  is typically calculated in the cold electron approximation for which $\partial \Delta \fkm / \partial_{k_{\gamma}} \ll \partial \fkm^0 / \partial_{k_{\gamma}}$ and is thus independent of the electric field. 

For sufficiently large fields that $\partial \Delta \fkm / \partial_{k_{\gamma}} \sim \partial \fkm^0 / \partial_{k_{\gamma}}$, the DC conductivity depends on the electric field and is defined with the relaxation operator:

\begin{equation}\label{general_ohmic_cond}
    \sigma_{\alpha \beta}(\mathcal{E}) =  \frac{2e^2}{k_B T \mathcal{V}_0} \sum_{m\kwv} v_{m \kwv, \alpha} (\Lambda_{m \kwv m' \kwv'}^{-1} \, v_{m'\kwv',\,\beta} \, f_{m'\kwv'}^0)
\end{equation}

Another important transport quantity, the  AC small-signal conductivity $\sigma_{\alpha \beta}^{\omega}$, describes the linear response of the system about a non-equilibrium steady-state \cite{Hartnagel_2001}. With the steady distribution $\fkm^s$ being set by a DC field $\mathcal{E}$ as described above, an AC field perturbation along crystal axis $\gamma$,
$\delta \mathcal{E}_{\gamma}(t) = \delta \mathcal{E}_{\gamma} e^{i \omega t}$,  induces a fluctuation of the steady distribution $\delta \fkm (t) = \delta \fkm (\omega) e^{i \omega t}$. This fluctuation is governed by the Fourier transformed Boltzmann equation:

\begin{equation}\label{ftacbte}
    \sum_{m'\kwv'}(i\omega \, \mathbb{I} + \Lambda)_{m\kwv m'\kwv'} \, \delta f_{m'\kwv'} = - \sum_{\gamma}\frac{e\delta \mathcal{E}_{\gamma}}{\hbar} \frac{\partial \fkm^s}{\partial_{k_{\gamma}}}
\end{equation}
Here, $\mathbb{I}$ is the identity matrix. The fluctuation in the distribution function induces a current fluctuation about the DC value, given as:

\begin{equation}\label{current_fluctuation}
    \delta j_{\alpha} = \frac{2 e}{\mathcal{V}_0} \sum_{m \kwv} v_{m \kwv,\,\alpha} \, \delta \fkm
\end{equation}

The small-signal AC conductivity is defined as the linear coefficient relating the current density variation to the perturbation, $\sigma_{\alpha\beta}^{\omega} \equiv \delta j_{\alpha}/\delta \mathcal{E}_{\beta} $. An explicit expression for AC conductivity can be obtained by combining the above expressions:

\begin{equation}\label{AC_conductivity}
    \sigma_{\alpha \beta}^{\omega} = \frac{2e^2}{\hbar \mathcal{V}_0} \sum_{m \kwv} v_{m \kwv,\alpha} \sum_{m'\kwv'} (i \omega \, \mathbb{I} + \Lambda)_{m\kwv m'\kwv'}^{-1} \, \bigg[-\frac{\partial f_{m'\kwv'}^s }{\partial_{k_{\beta}'}}\bigg]
\end{equation}

At equilibrium the steady distribution reduces to the equilibrium distribution $f_{m \kwv}^s = f_{m \kwv}^0$, the kinetic operator reduces to the scattering operator, $\Lambda_{m\kwv m'\kwv'} = \Theta_{m \kwv m' \kwv'}$. By examining Eq.~\ref{lowfield_cond} and Eq.~\ref{AC_conductivity}, we see that at equilibrium the zero-frequency differential conductivity is equal to the linear DC conductivity $ \sigma_{\alpha \beta}^{\omega = 0}(\mathcal{E} = 0) = \sigma_{\alpha \beta}^{lin}$.

\begin{subsection}{Fluctuations about a non-equilibrium steady state}

We now consider fluctuations about a non-equilibrium steady state induced by the stochastic nature of charge carrier scattering. Suppose that the steady state distribution $\fkm^s$ is known.  Just as in equilibrium, fluctuations in the instantaneous occupation of the quantum states occur. Microscopically, these fluctuations arise because of the stochastic nature of the scattering described by $\scm$. At steady state, detailed balance requires that the mean flux of particles into every quantum state is zero. However,  the flux of particles into or out of a quantum state is a Poissonian process and is characterized by a variance. Therefore, the instantaneous net flux into a quantum state is in general non-zero due to instantaneous imbalance between the incoming and outgoing fluxes \cite{GGK_1979}. Consequently, the occupancy of quantum states fluctuates under both equilibrium and non-equilibrium conditions. 

In the macroscopic limit at which fluctuations are observed in the laboratory, these distribution function fluctuations appear as instantaneous current fluctuations, or equivalently, as electronic noise. A non-random characteristic of these fluctuations is the spectral density of current density fluctuations, which, by the Wiener-Khintchine Theorem, is related to the single-sided Fourier transform of the autocorrelation of the current density fluctuations \cite{Hartnagel_2001}:

\begin{equation}\label{wiener_khintchine}
    S_{j_{\alpha}j_{\beta}}(\omega) \equiv (\delta j_{\alpha} \delta j_{\beta})_{\omega} = 2 \int_{-\infty}^{\infty} \overline{\delta j_{\alpha}(t) \delta j_{\beta}} e^{-i \omega t} dt
\end{equation}

\noindent
where the overbar indicates ensemble average. 

We seek to link the macroscopic current density fluctuations to microscopic distribution function fluctuations. Following Ref.~\cite{GGK_1979}, we now consider random fluctuations about the non-equilibrium steady state, $\delta \fkm(t) = \fkm(t) - \fkm^s$. In contrast to the fluctuations associated with the small signal conductivity, these fluctuations are induced by the stochastic nature of scattering rather than an external perturbation. The corresponding current density fluctuations can be expressed in terms of the fluctuation in the distribution function as in Eqn.~\ref{current_fluctuation}.

It follows that the ensemble average of the correlation function of instantaneous current fluctuations along axes $\alpha$ and $\beta$, $\overline{\delta j_{\alpha}(t) \delta j_{\beta}}$, can be expressed in terms of the correlation function of the  occupancy fluctuations, $\overline{\delta f_{m\mathbf{k}}(t) \delta f_{m_1\mathbf{k}_1}}$:

\begin{equation} \label{moment_correlation}
    \overline{\delta j_{\alpha}(t) \delta j_{\beta}} = \bigg( \frac{2 e}{\mathcal{V}_0}\bigg)^2 \sum_{m\mathbf{k}} \sum_{m_1\mathbf{k}_1} v_{m\kwv,\,\alpha} \, v_{m_1\kwv_1,\,\beta} \, \overline{\delta f_{m\mathbf{k}}(t) \delta f_{m_1\mathbf{k}_1}}
\end{equation}

Equation \ref{moment_correlation} shows that computing the spectral density of current density fluctuations requires calculating the correlations of single particle occupancy fluctuations $\overline{\delta f_{m\mathbf{k}}(t) \delta f_{m_1\mathbf{k}_1}}$. This function is known as the time-displaced, two particle correlation function \cite{GGK_1979}. Through a quantum statistical mechanical treatment, Gantsevich and coauthors have demonstrated that the time-displaced, two particle correlation function obeys the same Boltzmann equation as the fluctuation itself \cite{GGK_1969}: 

\begin{equation}\label{fluctuation_bte}
   \frac{\partial }{\partial t} \, \overline{\delta f_{m\mathbf{k}}(t) \delta f_{m_1\mathbf{k}_1}} + \sum_{m'\kwv'} \Lambda_{m\kwv m' \kwv'} \, \overline{\delta f_{m'\mathbf{k}'}(t) \delta f_{m_1\mathbf{k}_1}} = 0
\end{equation}

The result of Eqn.~\ref{fluctuation_bte} can also be justified less mathematically rigorously but with more physical intuition from Onsager's regression hypothesis (in particular, see Sec.~1 of Ref.~\cite{GGK_1979}).

Solving Eqn.~\ref{fluctuation_bte} requires specifying an initial condition, $\overline{\delta f_{m'\mathbf{k}'}\delta f_{m_1\mathbf{k}_1}}$, which is known as the one-time, two-particle correlation function. For a non-degenerate system with a fixed number of particles $N$, Fowler \cite{Fowler_1936} and Lax \cite{Lax_1960} derived the required condition as:

\begin{equation}\label{initial_condition}
    \overline{\delta f_{m\mathbf{k}}\delta f_{m_1\mathbf{k}_1}} = f_{m\mathbf{k}} \delta_{\mathbf{k} \mathbf{k}_1}\delta_{m m_1} - \frac{f_{m\mathbf{k}} f_{m_1\mathbf{k}_1}}{N}
\end{equation}

where the second term on the right-hand side indicates that a correlation exists between occupancies due to the fixed particle number. With this initial condition for the correlation, Eqns.~\ref{wiener_khintchine},  \ref{moment_correlation}, and \ref{fluctuation_bte} can be combined to express the spectral density of current fluctuations explicitly in terms of solutions to the Boltzmann equation. For a single band, we drop the band index to get:

\begin{equation}\label{current_correlation}
    (\delta j_{\alpha} \delta j_{\beta})_{\omega} = \bigg( \frac{2 e}{\mathcal{V}_0}\bigg)^2 \sum_{\mathbf{k}, \mathbf{k}_1} v_{\kwv,\alpha} \, v_{\kwv_1,\beta} \, (\delta f_{\kwv} \delta f_{ \kwv_1})_{\omega}
\end{equation}

\noindent As with the current density, the spectral density of distribution function fluctuations is related to its analagous correlation function by Fourier transform:

\begin{equation}\label{distribution_FT}
    (\delta f_{\kwv} \delta f_{\kwv_1})_{\omega} = \int_{-\infty}^{\infty} \overline{\delta f_{\kwv} (t) \delta f_{\kwv_1}} \, \text{e}^{-i \omega t} \, dt
\end{equation}

By exploiting the stationary property of the autocorrelation function,  the spectral density can be expressed as \cite{GGK_1979}:

\begin{equation}\label{split_HC}
    (\delta f_{\kwv} \delta f_{\kwv_1})_{\omega} = 2 \Re \bigg[\sum_{\kwv'}(i \omega \, \mathbb{I} + \Lambda)^{-1}_{\kwv \kwv'}\,  \overline{\delta f_{ \kwv'} \,\delta f_{\kwv_1}} \bigg]
\end{equation}

Combining Eqns.~\ref{initial_condition},  \ref{current_correlation}, and \ref{split_HC} we obtain the following expression:

\begin{align}\label{secondbte}
    S_{j_{\alpha}j_{\beta}}(\omega) &=
    2 \bigg(\frac{2 e}{\mathcal{V}_0}\bigg)^2 \Re \bigg[ \sum_{\mathbf{k}} v_{\kwv,\alpha} \sum_{\kwv'}  \, (i\omega \, \mathbb{I} + \Lambda)^{-1}_{\kwv \kwv'} \sum_{\mathbf{k}_1}\, v_{\kwv_1,\,\beta} \bigg(f_{\mathbf{k}'}^s \delta_{\mathbf{k}' \mathbf{k}_1} - \frac{f_{\mathbf{k}'}^s f_{\mathbf{k}_1}^s}{N}\bigg)\bigg] \nonumber \\
    &= 2 \bigg( \frac{2 e}{\mathcal{V}_0}\bigg)^2 \Re \bigg[\sum_{\kwv} v_{\kwv,\alpha} \sum_{\kwv'} (i\omega \, \mathbb{I} + \Lambda)^{-1}_{\kwv \kwv'} \bigg(f_{\mathbf{k}'}^s (v_{\kwv',\,\beta} - V_{\beta})\bigg)\bigg]
\end{align}

\noindent Here, $V_{\beta}$ is the drift velocity along the $\beta$ axis defined as:

\begin{equation}
    V_{\beta} = \frac{1}{N}\sum_{\kwv} v_{\kwv, \,\beta} f_{\kwv}^s
\end{equation}

\noindent where $N = \sum_{\kwv} f_{\kwv}$ is the total particle number.

From Eqn.~\ref{secondbte}, it follows that calculating the spectral density of the current fluctuations requires solving the inhomogeneous Boltzmann equation twice. First, the steady occupation function must be obtained using Eqn.~\ref{steady_bte}. Then, the Boltzmann equation is solved again with inhomogeneous term $f_{\kwv}^s (v_{\kwv,\beta} - V_{\beta})$ with $\fkm^s \equiv \fkm^{0} + \Delta \fkm$. The appropriate Brillouin zone integrations are then performed to calculate the power spectral density.

As a check of the above derivation, consider an equilibrium system for which $\mathcal{E}_{\gamma} = 0$ and $V_{\gamma} = 0$. The equation is simplified as $f_{\kwv}^s = f_{\kwv}^0$ and $\Lambda_{\kwv \kwv'} = \Theta_{\kwv \kwv'}$. Then, we have:

\begin{equation}
    S_{j_{\alpha}j_{\beta}}(\mathcal{E} = 0)  = 2 \bigg( \frac{2 e}{\mathcal{V}_0}\bigg)^2 \Re \bigg[\sum_{\kwv} v_{\kwv,\,\alpha}\sum_{\kwv'} (i\omega \, \mathbb{I} + \Theta)^{-1}_{\kwv \kwv'} \, f_{\mathbf{k}'}^0 \, v_{\kwv',\,\beta}\bigg] 
\end{equation}

With the same simplifications, the equilibrium AC conductivity from Eqn.~\ref{AC_conductivity} is:

\begin{equation}
    \sigma_{\alpha \beta}^{\omega}(\mathcal{E} = 0) = \frac{2e^2}{\hbar \mathcal{V}_0} \sum_{\kwv} v_{\kwv,\,\alpha} (i \omega \, \mathbb{I} + \Theta)_{\kwv \kwv'}^{-1} \, \bigg[-\frac{\partial f_{\kwv'}^0}{\partial_{k_{\beta}'}}\bigg]
\end{equation}

Combining the above expressions, we obtain the familiar Nyquist relationship \cite{Nyquist_1928}:

\begin{equation}\label{nyquist_noise}
    S_{j_{\alpha}j_{\beta}}(\mathcal{E} = 0)  = 4 \frac{k_B T_0}{\mathcal{V}_0} \, \Re [\sigma_{\alpha \beta}^{\omega}(\mathcal{E} = 0)]
\end{equation}

This relationship is formally valid only in equilibrium but remains approximately true in the `cold' electron regime for which $\partial \fkm^s / \partial_{\kwv} \approx \partial \fkm^0 / \partial_{\kwv}$  and $\Delta \fkm \ll \fkm^0$ such that  $\Lambda_{\kwv \kwv'} \approx \Theta_{\kwv \kwv'}$ and $\fkm^s \approx \fkm^0$.

\end{subsection}

\section{Numerical methods}\label{sec:Methods}
We now describe how to compute the spectral noise power and other quantities using the theory from the previous section. The inputs to the Boltzmann equation are the electronic structure and electron-phonon matrix elements $g_{m \kwv,m'\kwv + \qwv}$ calculated using electronic structure packages. First, the electronic structure and electron-phonon matrix elements for GaAs are computed on a coarse $8\times 8\times 8$ grid using density functional theory (DFT) and density functional perturbation theory (DFPT), respectively, with \textsc{Quantum Espresso} (QE) \cite{giannozzi_qe_2009, giannozzi_qe_2017}. These quantities are then interpolated to finer grids using Wannier interpolation with \textsc{Perturbo} \cite{perturbo_2020}. \textsc{Perturbo} includes corrections for polar materials that are necessary in GaAs \cite{mauri_2015_polarwannier}.

The electronic structure calculations using QE employ the same simulation parameters as in Ref.~\cite{jjzhou_2016}. Briefly, we use a plane wave cutoff of 72 Ryd and a relaxed lattice parameter of 5.556 $\textrm{\AA}$. We set the Fermi level to obtain a carrier concentration of 10$^{15}$ cm$^{-3}$ corresponding to a non-degenerate electron gas. We consider only conduction band electronic states within an energy cutoff of 335 meV above the conduction band minimum (CBM). This energy window is larger than the window used in typical electron transport calculations since the present calculations allow the electric field to heat the electrons, leading to occupation at higher energy states. Further increasing the energy window by 50 meV had negligible effect on the observables of interest like spectral noise power.

In \textsc{Perturbo}, we use a grid of $200\times 200\times 200$ for the Wannier interpolated electronic structures and electron-phonon matrix elements. The transition rates of Eqn. \ref{matrixel} are calculated at 300 K. We consider convergence by determining the change in the spectral noise power at the maximum electric field for which the $\Delta \fkm \ll \fkm^0$ assumption is satisfied. Numerical experimentation shows that this condition is satisfied for $\mathcal{E} \lesssim  800$ \vcm. The spectral noise power at 800 \vcm \, using the $200\times 200\times 200$ grid differs by less than 1\% from the value obtained on a grid with twice the number of grid points. The delta function in the electron-phonon matrix elements is approximated with a Gaussian with a 10 meV broadening parameter \cite{jjzhou_2016}. Decreasing the broadening to 6 meV changed the spectral noise power at 800 \vcm \, by less than 3\%.

While \textsc{Perturbo} performs the Wannier interpolation for the electron-phonon interaction on fine grids \cite{Pizzi_2020}, it does not explicitly construct the collision matrix of Eqn.~\ref{matrixel}. Instead, the mobility is computed using an iterative scheme under the cold electron approximation \cite{Bernardi_2016}. We found that this iterative method was numerically unstable for the warm electron case. We instead solved the linear system using the Generalized Minimal Residual (GMRES) algorithm as implemented in the Scientific Python library \cite{scipy-2020}. The matrix was constructed by modifying \textsc{Perturbo} to output the elements of Eqn.~\ref{matrixel}.

As described in Sec. \ref{sec:Theory}, the derivative term corresponding to particle drift in an electric field $\mathcal{E}$ is approximated by a finite difference matrix. Boundary conditions must be applied to points that do not have a full set of first-nearest neighbors. To do so, we assume that these points have zero occupation by removing the contributions of these states to the finite difference matrix. The energy window is selected so that these boundary states indeed have negligible population, also ensuring that scattering induced by the collision matrix $\scm$ for these states can be neglected. 

With the collision and drift matrices computed, we then construct the relaxation operator $\Lambda_{m\mathbf{k}m'\mathbf{k'}}$, Eqn.~\ref{lambda}.  The steady-state distribution is obtained by solving the resulting linear system given by Eqn.~\ref{steady_bte}. We then solve Eqn.~\ref{secondbte} with the inhomogenous term constructed from the previously computed steady-state distribution as input. For this second Boltzmann equation, we include an $i\omega$ term on the diagonal of the linear system which corresponds to the Fourier transformed time derivative. Finally, the spectral density is computed as a Brillouin zone integration over the distribution that solves the second Boltzmann equation. The calculation of the AC mobility proceeds in a similar way as for the steady distribution except with the addition of the $i\omega$ term.

In this work, we apply our method to GaAs. Owing to computational limitations, we account for first-order electron-phonon (1ph) processes and  neglect higher-order interactions that are reported to play a role in GaAs \cite{Lee_2018_b}.  Also, recently studies report the effect of quadrupole electron-phonon interactions on electron transport \cite{jhalani_quad_2020,brunin_quad_2020}. In particular, the work of Ref.~\cite{brunin_quad_2020}  predicted a significant correction to the mobility in GaAs limited by acoustic mode scattering. Our  calculations were performed at 300 K at which the scattering is dominated by polar optical phonons, and so we neglected quadrupole interactions. Finally, we note that the method described above is easily extendable to other technologically interesting semiconductors. In particular, first-principles calculations of electron-phonon interactions in Si, InP, and Ge are now routine; calculations in these materials is the subject of ongoing work.

\section{Results}\label{sec:Results}

\begin{figure*}
\includegraphics[width=\textwidth]{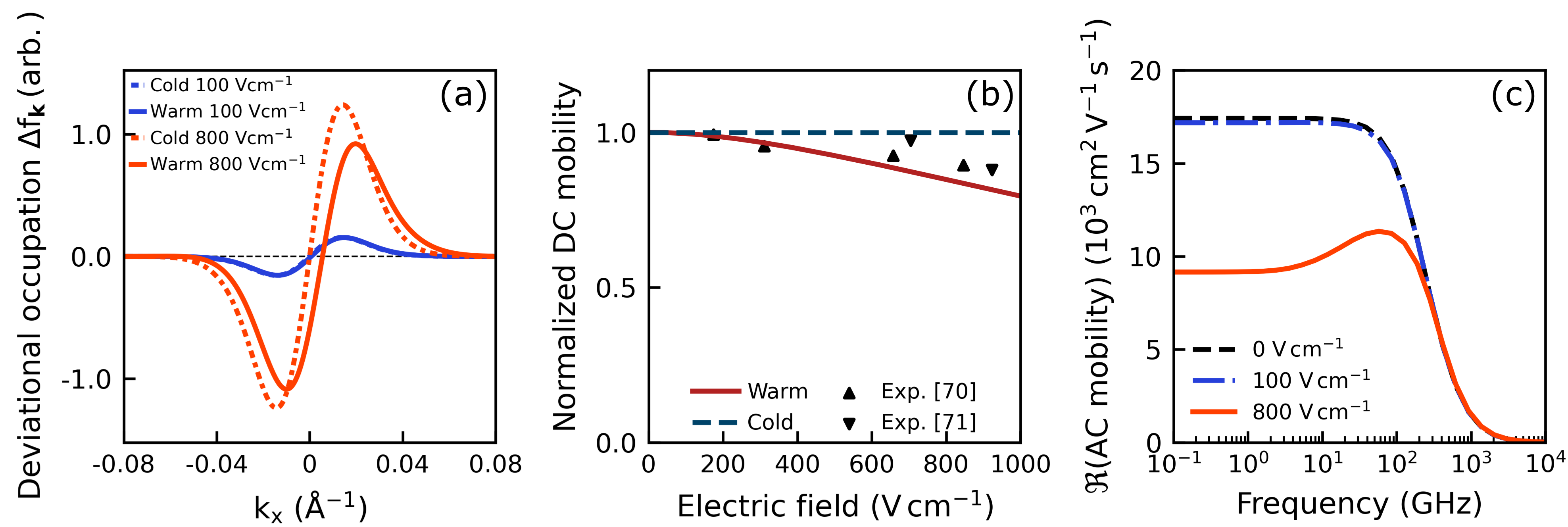}
\caption{ (a) Deviational occupation $\Delta f_k$ in GaAs at 300 K under the cold (dotted lines) and warm (solid lines) electron approximations versus longitudinal wave vector $k_{x}$. Curves plotted for $\mathcal{E} = 100 \, \rm V \, cm^{-1}$ (blue), and $\mathcal{E} = 800 \, \rm V \, cm^{-1}$ (orange). The dashed black line is a guide to the eye. (b) Normalized longitudinal ($\parallel$) DC mobility versus electric field of the cold (dashed  blue line) and warm electrons (solid red line). The heating of the electrons leads to a decreased mobility. The trend of the normalized mobility agrees well with experiments: Figure 1, Ref.\,\cite{Bareikis_1986} (Upward black triangles) and Figure 4, Ref.\,\cite{Kino_1968} (Downward black triangles). (c) Real part of the longitudinal small-signal AC mobility versus frequency for equilibrium (dashed black line),  $\mathcal{E} = 100 \, \rm V \, cm^{-1}$ (dash-dot blue line), and $\mathcal{E} = 800 \, \rm V \, cm^{-1}$ (solid orange line) under the warm electron approximation. The AC mobility exhibits spectral features at frequencies that are characteristic of the inverse momentum and energy relaxation times (see Section \ref{subsec:freqdep}).}
\label{fig:S1}
\end{figure*}

\subsection{Transport}

We begin by examining the steady state distribution and associated transport observables in the cold and warm electron regimes. Figure~\ref{fig:S1}a plots the deviational steady state distribution functions under the two approximations versus wave vector parallel to the electric field, $k_x$. We  refer to this direction as the longitudinal direction. At low fields $\mathcal{E} < 100 \, \rm V \, cm^{-1}$, the solutions are nearly identical, but as the field increases, differences in the distribution functions emerge. Under the cold electron approximation, Eqn.~\ref{steady_bte} shows that $\Delta \fkm$ is required to possess odd symmetry about the Brillouin zone center because  $\partial \fkm^0 / \partial_{\kwv}$ is odd with respect to $k_x$ while the scattering matrix is even ($\Theta_{\kwv \kwv'} = \Theta_{-\kwv -\kwv'}$); this symmetry is evident in Fig.~\ref{fig:S1}a. In contrast, in the warm electron case the electrons can be heated and the solution becomes asymmetric with increasing field. 

The transport properties of the warm electron distribution differ from those of the cold distribution because warm electrons in the high energy tail are able to emit optical phonons and hence exhibit higher scattering rates. As reported previously \cite{jjzhou_2016}, the predicted mobility of GaAs exceeds the experimental mobility owing to the exclusion of higher-order phonon scattering processes and the  lower calculated effective mass (0.055 $m_0$) compared to experiment (0.067 $m_0$) \cite{Lee_2018_b}. 

Therefore, to facilitate comparison we examine the  DC mobility normalized by its low-field value in Fig.~\ref{fig:S1}b. The low-field value of the computed mobility is 17,420 $\rm cm^2 \, V^{-1} \, s^{-1}$.  At low fields $\mathcal{E} < 100 \, \rm V \, cm^{-1}$, the mobility under the warm and cold electron approximations agrees to within 1$\%$. At $\mathcal{E} = 800 \, \rm V \, cm^{-1}$, the DC mobility of the warm electrons has decreased by more than $10\%$. This behavior is qualitatively consistent with the sublinear current voltage characteristic (CVC) of n-type GaAs \cite{Hartnagel_2001}, or a decrease in mobility with increasing electron temperature. The field dependence of the normalized mobility shows favorable comparison to experiment, implying that our calculation is properly capturing the heating with the field.

In addition to steady quantities, the small-signal AC mobility can be computed as in Eqn.~\ref{AC_conductivity}. Figure~\ref{fig:S1}c presents the small-signal AC mobility for the warm electron gas versus frequency for several electric fields. At zero frequency, the equilibrium AC mobility is equal to the equilibrium DC mobility, as expected. The decrease of the AC mobility with electric field at is also consistent with the trend observed in the DC mobility. At $f\sim 1$ THz, the AC field frequency exceeds the phonon-mediated scattering rates which redistribute the electrons, and thus the AC mobility rolls off at all fields. This result reflects the electrical response transitioning from a purely resistive to a purely reactive regime as the frequency exceeds the highest scattering rates.

The frequency dependence of the AC mobility indicates the relevant timescales of momentum and energy relaxation \cite{Teitel_1982}. In particular, for 800 \vcm, we observe a lower value of the AC mobility at low frequency, followed by a maximum at around $100$ GHz.  This feature is due to energy exchange with phonons and will be discussed in Section \ref{subsec:freqdep}.

\subsection{Diffusion noise}\label{subsec:Static noise}

\begin{figure*}
\includegraphics[width=\textwidth]{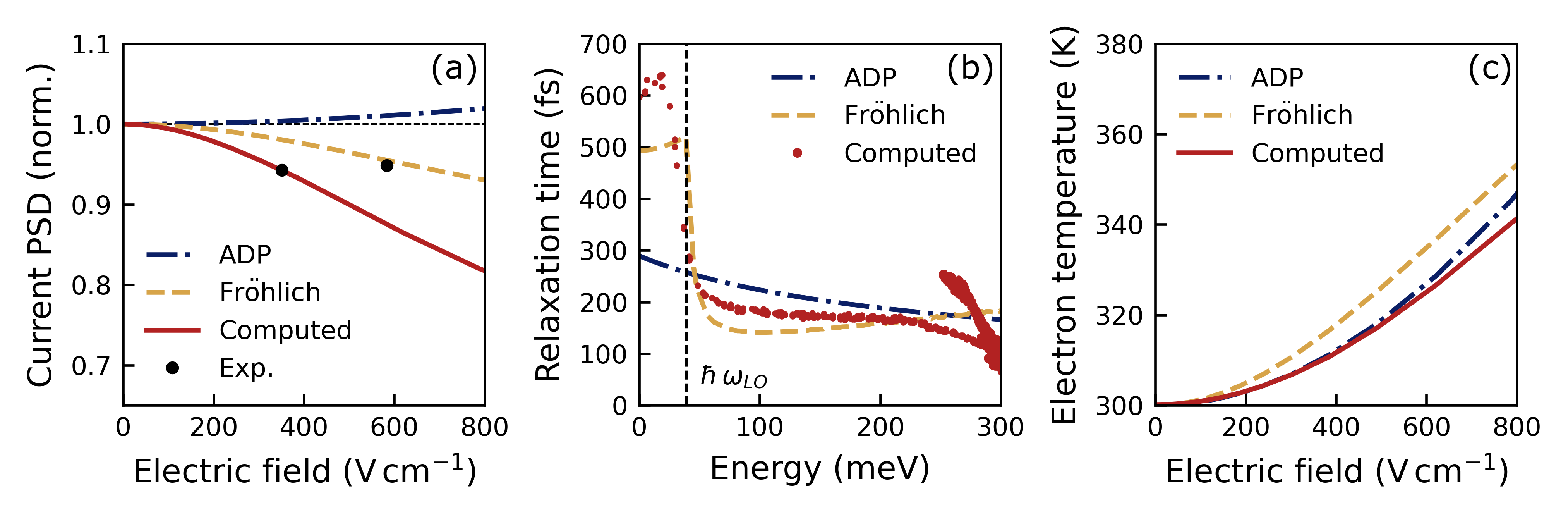}
\caption{(a) Spectral density of  longitudinal current density fluctuations  (solid red line) normalized to the Nyquist value versus electric field along with Davydov spectral densities calculated using ADP (dash-dot blue line) and  Fr\"{o}hlich (dashed yellow line). At equilibrium, the noise agrees with Nyquist-Johnson noise (dotted black line). The \textit{ab-initio} calculation predicts a steeper decrease in current PSD with field compared to the approximations. The symbols correspond to experimental measurements (Figure 11, Ref.~\cite{Bareikis_1994}). (b) Relaxation time versus energy above conduction band minimum for GaAs at 300 K using ADP (dash-dot blue line), Fr\"{o}hlich potential (dashed yellow line), and computed (red circles). The energy of the zone-center LO phonon is shown for reference (dashed black line). (c) Effective electron gas temperature versus electric field  for ADP (dash-dot blue line), Fr\"{o}hlich (dashed yellow line), and computed (solid red line). The magnitude of electron heating is similar among the various calculations.}
\label{fig:S2}
\end{figure*}

We now calculate the spectral density of current fluctuations from the non-equilibrium steady state in GaAs. Figure~\ref{fig:S2}a shows the spectral density of longitudinal current fluctuations versus electric field at an observation frequency of 1 MHz, far smaller than any scattering rate. At equilibrium, the noise is given by the Nyquist relation, Eqn.~\ref{nyquist_noise}. It is conventional to report the spectral density normalized to the Nyqist value to allow comparison between samples of different carrier density \cite{Bareikis_book}.

As the electric field increases, the computed noise decreases below the Nyquist value. Few experimental studies of noise in GaAs cover the fields of present interest, but reasonable  agreement is observed with measurements by Bareikis \textit{et al.} \cite{Bareikis_1994}. We note that a decrease with field is observed in other studies in GaAs \cite{Bareikis_1986,Bareikis_1980} though the sparsity of data in the relevant electric field range prevents direct comparison. 

To better understand the decreasing trend, we use an approximate solution of the Boltzmann equation for an electron gas interacting quasi-elastically with a thermal phonon bath \cite{Kogan_1996, conwell_1968}. Under the quasi-elastic approximation, the distribution function is expanded in momentum space using Legendre polynomials. Because the distribution is nearly isotropic in momentum space under quasi-elastic scattering, only the two lowest Legendre polynomials need be retained \cite{Davydov_1937}; the zeroth-order term gives the occupancy versus energy and is known as the Davydov distribution. The model is parametrized by the energy dependence of the momentum and energy relaxation times, $\tau$ and $\tau_{\epsilon}$ respecitvely, and the  inelasticity ratio $\tau / \tau_{\epsilon}$ \cite{Hartnagel_2001}. Once these parameters are specified,  the Davydov distribution can be computed and used with Eqn.~\ref{secondbte} to calculate the spectral density of current fluctuations \cite{GGK_1979}.

Approximate analytic expressions for the electron relaxation times in semiconductors are available \cite{Lundstrom_2000}.  Previous works have calculated the Davydov distribution for a power-law energy dependence of the relaxation times such as that from the acoustic deformation potential (ADP) \cite{Skullerud_1969,Katilius_1966,SCHLUP1973485}. However, in GaAs at room temperature, the long-ranged Fr\"{o}hlich interaction with longitudinal optical (LO) phonons is the dominant scattering mechanism \cite{jjzhou_2016,Liu_2017}. 

\begin{figure*}[!htb]
\includegraphics[width=\textwidth]{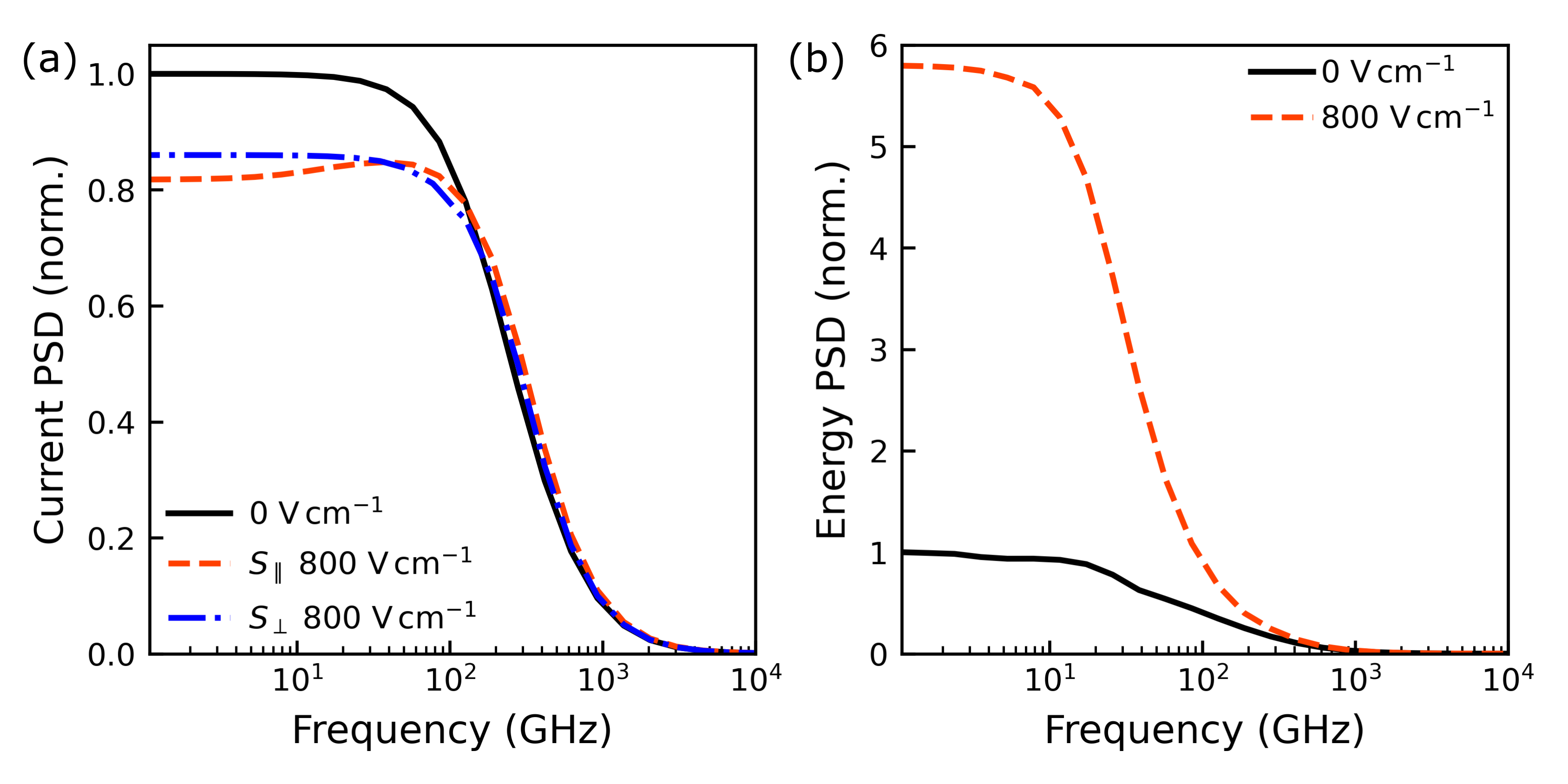}
\caption{(a) Computed power spectral density (PSD) of longitudinal ($\parallel $, dashed orange line) and transverse ($\perp $, dashed-dotted blue line) current density fluctuations versus frequency at $\mathcal{E} = 800$ $\rm V \, cm^{-1}$, along with the Nyquist-Johnson prediction for $\mathcal{E}=0$ (solid black line). (b) Spectral density of energy fluctuations versus frequency at equilibrium (solid black line), $\mathcal{E} = 800 \, \rm V \, cm^{-1}$ (dashed orange line). The time scale for electron temperature fluctuations sets the upper frequency limit for the convective mechanism.}
\label{fig:S3}
\end{figure*}

In Fig.~\ref{fig:S2}a, we compare the \textit{ab-initio} longitudinal spectral density to that predicted using the Davydov distribution with the ADP and Fr\"{o}hlich scattering rates. The approximate relaxation times have been scaled to match the computed low-field  mobility, and the inelasticity ratio has been selected using an estimation of the energy and momentum relaxation times (see Fig.~\ref{fig:S3}). The spectral density is observed to decrease monotonically with the electric field. This decrease is captured qualitatively by the Fr\"{o}hlich calculation.  In contrast, the ADP noise increases monotonically with field.

These trends can be understood in terms of the differing energy dependencies of the relaxation times in the various approximations. Figure~\ref{fig:S2}b shows the phonon-mediated relaxation times versus energy for electrons in GaAs at 300 K for the three cases. Below the zone-center LO phonon energy $\hbar \omega_{LO} \sim 35$ meV, the computed relaxation times are set by LO phonon absorption \cite{Zhou_2016}. Above the LO phonon energy, LO emission becomes dominant and the relaxation times sharply decrease to a value that remains roughly constant until electron energies are near the L-valley minimum at $\sim 0.25$ eV above the CBM. This absorption-to-emission transition is qualitatively captured by the Fr\"{o}hlich approximation. The ADP relaxation times agree reasonably well with the computed ones in the emission-dominated region but do not exhibit the absorption-to-emission transition.

The electric field dependence of the spectral noise power reflects the balance between the growth of scattering rates with electron energy and the heating of the electron gas by the DC field \cite{ANINKEVICIUS_1993}. To understand this balance, we examine the effective electron temperature of the steady distribution for the three cases in Figure~\ref{fig:S2}c. The effective electron temperature is calculated as the temperature of a Maxwell-Boltzmann distribution that yields the same  energy density as the steady non-equilibrium distribution. At low fields $\mathcal{E} < 100$ \vcm, the temperature is equal to the lattice temperature. As the electric field increases, the effective temperature increases, corresponding to occupancy at higher energies and increased scattering rates. Near equilibrium where the mobilities are equivalent, the temperature rise predicted from each approximation is similar, but at higher fields, the \textit{ab-initio} calculation predicts a slightly lower temperature than do either the ADP or Fr\"{o}hlich approximations.

As the electron gas heats, higher energy states are occupied and thus the spectral noise power, Eqn.~\ref{secondbte}, includes contributions from fluctuations in those states; hence, the spectral noise power may increase on heating. On the other hand, at these high energies, the scattering events which damp out fluctuations are more frequent, tending to decrease the noise. The competition between these mechanisms sets the trends shown in Figure~\ref{fig:S2}a. For both Fr\"{o}hlich and the present calculations, the sharp increase in scattering rates associated with the absorption-to-emission transition dominates, and the spectral density decreases monotonically with electric field. In contrast, the ADP approximation shows increasing noise with electric field as the heating of the electrons dominates the weak increase of the scattering rates.

The evolution of the spectral density with electric field demonstrates the sensitivity of the spectral noise power to the energy dependence of the scattering rates. Although the mobility at equilibrium is equivalent for all three cases, the non-equilibrium noise behavior exhibits qualitatively different trends depending on the energy dependence and inelasticity of the scattering mechanisms.

\subsection{Spectral noise power}\label{subsec:freqdep}

The non-equilibrium noise exhibits spectral features that are not present in the Nyquist-Johnson case. Figure~\ref{fig:S3}a shows the spectral density of  longitudinal (L) and transverse (T) current fluctuations (relative to the electric field axis) versus frequency at $\mathcal{E}=800$ $\rm V\, cm^{-1}$.  There are several notable features of the spectral density in this figure. First, the spectral density is constant at low frequencies and rolls off as frequency increases, decreasing to 50\% of its low frequency value at 300 GHz. Secondly, an anisotropy exists  between the longitudinal and transverse spectral densities. Finally, the longitudinal noise exhibits a non-monotonic trend for frequencies around 50 GHz, similar to that observed for the AC mobility in Fig.~\ref{fig:S1}c. Spectroscopic measurements of the noise power at these frequencies have not been performed, but these trends are qualitatively similar to those observed in recent Monte Carlo simulations \cite{mateos_2017}.

We discuss each of these points in turn. Consider first the noise at equilibrium. The zero-field curve shows that the longitudinal and transverse spectral densities are equal and coincide with the Nyquist-Johnson value, Eqn.~\ref{nyquist_noise}. As with the AC mobility, the spectral density rolls off at frequencies exceeding the phonon-mediated scattering rates because the electronic system cannot redistribute in response to the fluctuation. This roll-off behavior has been noted previously \cite{conwell_1967} and has also been observed for phonon thermal conductivity (see Fig.~1b in Ref.~\cite{Chatput_2013}).

Now consider the noise with $\mathcal{E} = 800$ $\rm V \, cm^{-1}$.  A similar roll-off with increasing frequency as the equilibrium case is observed. At low frequency, both the longitudinal and transverse spectral densities are lower than the Nyquist value because of the increased electron temperature. However, an anisotropy exists in the spectral densities. The origin of this feature is the `convective' mechanism \cite{Kogan_1996,Price_1965,Hartnagel_2001} and can be understood by decomposing the current fluctuations into two sources. The first is the fluctuation of the drift velocity, induced by stochastic transitions between states of differing group velocity. The second is the fluctuation of the electron temperature, induced by random energy exchange with the thermal phonon bath. Under non-equilibrium conditions, these fluctuations couple. As the gas is heated by the electric field, the fluctuating current induces a variation in the Joule heating. The resulting electron temperature fluctuation changes the conductivity, which in turn modifies the current. This coupling only exists for fluctuations longitudinal to the electric field because transverse fluctuations do not affect Joule heating. In sublinear CVC materials such as GaAs, the conductivity decreases with electron temperature, and the convective mechanism suppresses longitudinal fluctuations. This feature is indeed observed in Fig.~\ref{fig:S3}a.

The convective mechanism is only present at frequencies $\omega\tau_{\epsilon} \ll 1$, where $\tau_{\epsilon}$ is the energy relaxation time. As discussed above, the local maxima from the convective contribution appears at $\omega \tau_{\epsilon} = 1$ in the longitudinal direction (see Ref.~\cite{Hartnagel_2001}, Chapter 7).  The energy relaxation time can also be extracted by calculating the spectral density of electron temperature fluctuations versus frequency. This calculation is the energy analogue of Eqn.~\ref{secondbte}, where the relevant state quantity is the energy instead of the group velocity. Figure \ref{fig:S3}b shows the spectral density of energy fluctuations versus frequency for several electric fields. At low frequencies, $f<10$ GHz, the spectral density increases with field as the temperature fluctuations rise with higher Joule heating. At higher frequencies, $f \sim 50$ GHz, the energy fluctuations decrease to 50\% of their low frequency values and begin to converge for the two fields shown. This convergence signifies that the temperature of the electron gas cannot change sufficiently rapidly due to its finite thermal capacitance. Consequently, the convective noise mechanism is removed and the anisotropy of the densities in Fig.~\ref{fig:S3}a also disappears; the longitudinal and transverse spectral densities converge. The convective mechanism  is also responsible for the non-monotonic trend of the AC mobility seen in Fig.~\ref{fig:S1}c.

\subsection{Quasi-elastic scattering}

\begin{figure*}[!htb]
\includegraphics[width=\textwidth]{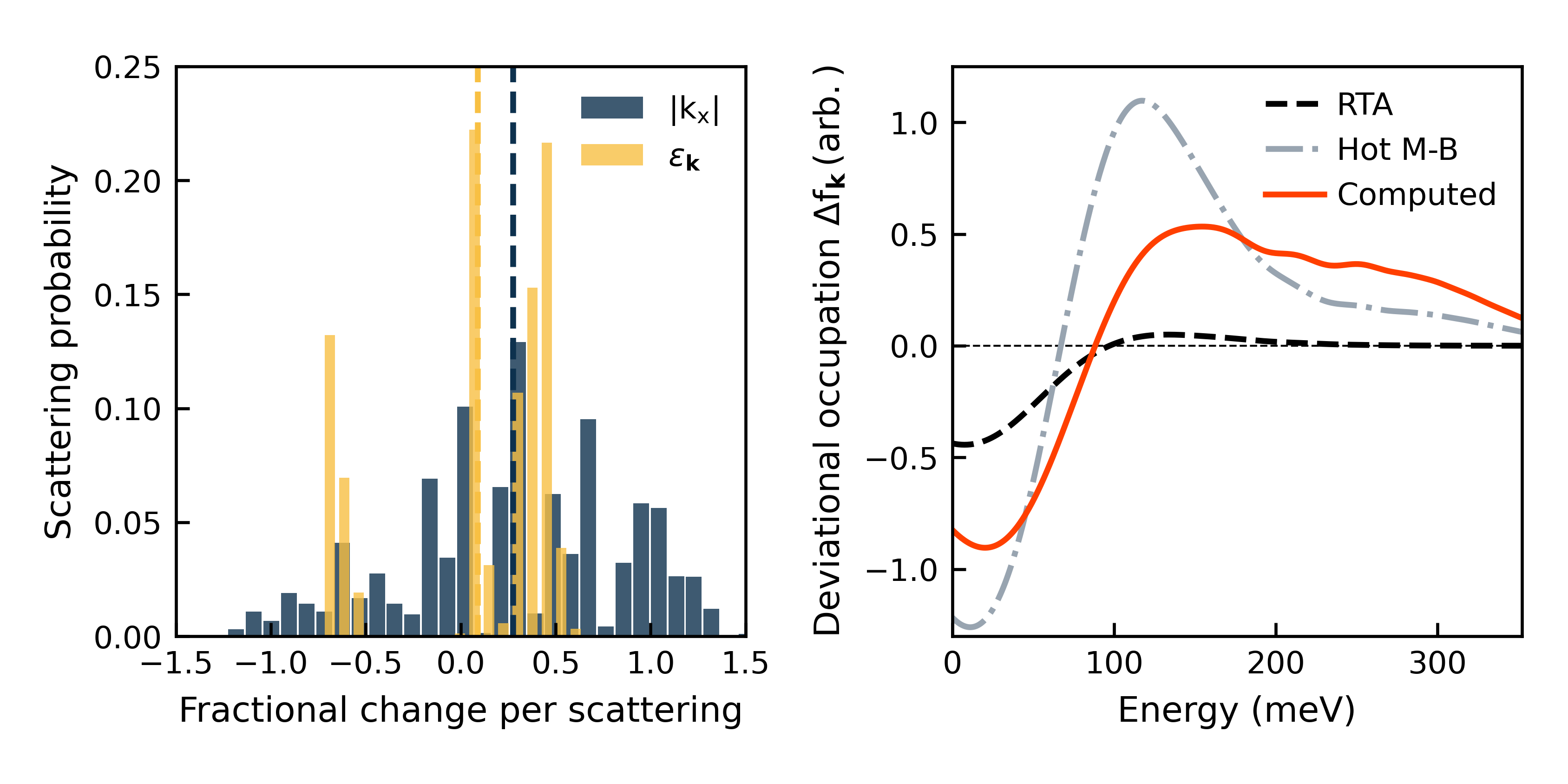}
\caption{(a) Probability histograms of  longitudinal momentum loss $R_{\kwv}$ (blue bars) and  energy loss $R_{\epsilon_{\kwv} }$ (yellow bars) normalized by the thermal averages at 800 $\rm V \, cm^{-1}$. The dashed lines represent the average transfer per scattering event. At this field, the average fractional dissipation of longitudinal momentum is $\sim 3\times$ larger than that for energy. (b) Deviational occupation $\Delta f_{\kwv}$ in GaAs at 300 K versus  energy calculated under the RTA (dashed black line), hot Maxwell-Boltzmann (dashed-dotted grey line), and \textit{ab-initio} warm electron approximation (solid orange line) at 800 $\rm V \, cm^{-1}$. The dashed black line is added as a guide to the eye. Neither the RTA nor the Maxwell-Boltzmann capture the hot electron tail.}
\label{fig:S4}
\end{figure*}

The present formalism for electronic noise permits the study of the microscopic processes responsible for electronic noise in a manner that is difficult to obtain by other methods. As an example, consider the spectral features present in Fig.~\ref{fig:S3}. Comparing the frequency where the current power spectral density and energy power spectral density reach half of their low frequency values (300 GHz versus 50 GHz, respectively), the energy relaxation time is inferred to be around 6 times longer than the momentum relaxation time, implying that the quasi-elastic assumption is valid. This observation is surprising given the well-known dominance of high-energy LO phonon emission in GaAs \cite{Zhou_2016} and that inelasticity is expected only when the physical temperature is comparable to the Debye temperature \cite{conwell_1967}. Analytical treatments of noise under dominant LO phonon coupling typically assume strongly inelastic interactions between the electrons and lattice (see Sec. 3.8 of Ref.~\cite{Kogan_1996}, Sec. 7.3 of Ref.~\cite{Hartnagel_2001}, or Ref.~\cite{Rabinovich_1969}).

We identify the origin of this discrepancy by examining how individual scattering events contribute to the momentum and energy relaxation of the electron system to the phonons. These transfers can be expressed as sums over each of the electron-phonon scattering processes in the collision integral weighted by the energy and momentum of the mediating phonon. Every electronic state in the BZ is coupled via phonons to other states; by summing over all possible scattering processes, we obtain the average energy and momentum exchanged in a single scattering event. More precisely, the fractional change in momentum and energy per scattering event are calculated from:

\begin{equation}
    R_{x} = \frac{1}{\Theta_{\kwv\kwv}\langle|x|\rangle} \sum_{\kwv'}\Delta x \, \Theta_{\kwv'\kwv}
\end{equation}

where $x = k_x, \epsilon_k$ and $\Delta x = x-x'$ is the difference in the state quantity between $\kwv$ and $\kwv'$. $\langle|x|\rangle$ denotes the thermal average magnitude of the relevant quantity; $\Theta_{\kwv'\kwv}$ represents the component of the diagonal element of the collision matrix (the scattering rate) corresponding to scattering from $\kwv$ to $\kwv'$; and other variables carry same meaning as defined in Section \ref{sec:Theory}.

These fractional changes at 800 $\rm V \, cm^{-1}$  are plotted as a probability histogram in Figure ~\ref{fig:S4}a. In this figure, we have binned each state in the BZ by the value of $R_{\mathbf{k}}$ and $R_{\epsilon_{\mathbf{k}}}$. For all the states in a given bin, we calculate  the probability of scattering $\mathcal{P} \propto \sum_{\textrm{bin}} \Theta_{\kwv \kwv} \fkm^s$ (the final quantity is normalized to unity). The horizontal position indicates the average fractional change in energy or momentum induced by the event. Positive values of the fractional change correspond to net transfers to the lattice, or dissipation, while negative values correspond to transfers to the electrons, or accumulation. The height of a  bar represents the population-weighted probability of scattering in a given time interval.

Figure ~\ref{fig:S4}a reveals several important features. First, energy transfers are clustered into two groups. The  grouping of  accumulation events around $-0.75$ corresponds to the $\sim 35 \, \rm meV$ energy gain associated with LO absorption, which dominates scattering of electrons below the emission threshold $\hbar \omega_{LO}$. The relatively disperse grouping of the dissipation events reflects a balance between LO emission and absorption for states above the threshold. Second, in contrast to energy transfers, momentum transfers grow with the wave vector of the mediating phonon. Consequently, a broader and more disperse distribution of momentum transfers is available. Finally, the balance between dissipation and accumulation differs between energy and momentum. In equilibrium, these processes are balanced, but at 800 $\rm V \, cm^{-1}$, the net transfers for both quantities are dissipative as the warm electrons transfer excess momentum and energy to the lattice. The dashed lines in the figure represent the average fractional transfer per scattering event and indicate that the net momentum dissipation exceeds the energy dissipation by around a factor of 3. This imbalance is partly responsible for the disparate time scales of energy and momentum relaxation observed in Fig.~\ref{fig:S3}.

The second contributing factor to the relatively long energy relaxation time is the presence of a hot electron tail in the calculated distribution.  In Fig.~\ref{fig:S4}b, we plot the steady deviation distribution, $\Delta f_s$, calculated under the warm electron approximation using the full \textit{e}-ph scattering matrix versus energy. For reference, the corresponding distributions for a hot Maxwell-Boltzmann at the non-equilibrium electron temperature and a `relaxation-time distribution' obtained under the warm electron approximation with only the on-diagonal elements of the scattering matrix. The \textit{ab-initio} treatment predicts a hot electron tail that is not observed with either approximate method. Although representing only a small fraction of the population, these hot electrons are at energies $5-10 \times$ the thermal average value. Consequently, many scattering events are needed to return these electrons to equilibrium, further increasing the energy relaxation time. The result is that the quasi-elastic approximation is unexpectedly accurate despite the inelastic nature of optical phonon scattering, and thus explaining the features in the spectral noise power and AC mobility.

\section{Summary and Future Outlook}

The primary numerical tools used to study electronic noise are Monte Carlo (MC) methods \cite{Rodilla_2013, mateos_2015, Mateos_2008,Mateos_2009, mateos_2017}. These simulators have many advantages, including the ability to incorporate realistic device geometries and space charge effects through coupled Poisson solvers, and they are thus useful to interpret experimental measurements on devices. However, MC studies rely upon semi-empirical models of scattering and electronic structure that require parameters such as deformation potentials, sound velocities, effective masses, and energy gaps to be specified and calibrated against experiment. The methods are thus most useful for well-characterized materials for which these empirical models are available.

A parallel development in the transport field has been the introduction of \textit{ab-initio} methods to study low-field transport phenomena  without adjustable parameters. These methods enable not only the computation of low-field transport properties such as electronic mobility \cite{Giustino_2017,Bernardi_2016} and phonon thermal conductivity \cite{lindsay_phon_survey} but also an understanding of the microscopic scattering processes that underlie these macroscopic properties and prediction of the properties of new materials. However, thus far these methods have been restricted to the low-field, cold electron regime.

In this work, we have described an \textit{ab-initio} theory of electronic noise for warm electrons in semiconductors. The method requires no adjustable parameters, with the phonon dispersion, band structure, and electron-phonon coupling calculated from first-principles. Further, this method permits the study of transport even when the electrons are not in equilibrium with the lattice, being free of the cold electron approximation used in previous transport studies. To demonstrate the method, we performed calculations in GaAs, a technologically relevant material, and demonstrated that the spectral features of the AC mobility and current noise are linked to the disparate time scales of energy and momentum relaxation. The quasi-elastic approximation is unexpectedly accurate in GaAs despite the dominance of polar optical phonon scattering.  Our work paves the way for  first-principles studies of electronic noise in other semiconductors that will advance the study of transport phenomena and applications of low-noise semiconductor devices.

\begin{acknowledgements}
The authors thank Jin-Jian Zhou, I-Te Lu, Vatsal Jhalani, and Marco Bernardi for assistance with \textsc{Perturbo} and useful discussions. This work was supported by AFOSR under Grant Number FA9550-19-1-0321. AYC  acknowledges the support of the National Science Foundation Graduate Research Fellowship.
\end{acknowledgements}

\bibliographystyle{is-unsrt}
\bibliography{references}{}

\end{document}